\documentclass[aps,twocolumn,showpacs,superscriptaddress]{revtex4-1}
\usepackage{graphicx}
\usepackage{amsmath}
\usepackage{amssymb}
\usepackage{bm}
\usepackage{multirow}
\usepackage{appendix}
\usepackage{color}
\usepackage{placeins}

\usepackage[colorlinks,
 linkcolor=blue,
 anchorcolor=blue,
 citecolor=blue,
 urlcolor=blue
 ]{hyperref}

\begin{document}
\title{An $O(N)$ $Ab~initio$ Calculation Scheme for Large-Scale Moir\'{e} Structures}
\author{Tan Zhang}
\affiliation{
Beijing National Laboratory for Condensed Matter Physics, and Institute of Physics, Chinese Academy of Sciences, Beijing 100190, China
}
\author{Nicolas Regnault}
\thanks{regnaultn@gmail.com}
\affiliation{
Department of Physics,
Princeton University,
Princeton, NJ 08544, USA 
}

\affiliation{
Laboratoire de Physique de l'Ecole normale sup\'{e}rieure, ENS, Universit\'{e} PSL, CNRS, Sorbonne Universit\'{e}, Universit\'{e} Paris-Diderot, Sorbonne Paris Cit\'{e}, Paris, France
}

\author{B. Andrei Bernevig}
\affiliation{
Department of Physics,
Princeton University,
Princeton, NJ 08544, USA
}

\author{Xi Dai}
\affiliation{
Department of Physics, Hong Kong University of Science and Technology, Kowloon, Hong Kong
}

\author{Hongming Weng}
\thanks{hmweng@iphy.ac.cn}
\affiliation{
Beijing National Laboratory for Condensed Matter Physics, and Institute of Physics, Chinese Academy of Sciences, Beijing 100190, China
}
\affiliation{
School of Physical Sciences, University of Chinese Academy of Sciences, Beijing 100049, China
}
\affiliation{
Songshan Lake Materials Laboratory, Dongguan, Guangdong 523808, China
}

%\date{\today}

\begin{abstract}
We present a two-step method specifically tailored for band structure calculation of the small-angle moir\'{e}-pattern materials which contain tens of thousands of atoms in a unit cell. In the first step, the self-consistent field calculation for ground state is performed with $O(N)$ Krylov subspace method implemented in OpenMX. Secondly, the crystal momentum dependent Bloch Hamiltonian and overlap matrix are constructed from the results obtained in the first step and only a small number of eigenvalues near the Fermi energy are solved with shift-invert and Lanczos techniques. By systematically tuning two key parameters, the cutoff radius for electron hopping interaction and the dimension of Krylov subspace, we obtained the band structures for both rigid and corrugated twisted bilayer graphene structures at the first magic angle ($\theta=1.08^\circ$) and other three larger ones with satisfied accuracy on affordable costs. The band structures are in good agreement with those from tight binding models, continuum models, plane-wave pseudo-potential based $ab~initio$ calculations, and the experimental observations. This efficient two-step method is to play a crucial role in other twisted two-dimensional materials, where the band structures are much more complex than graphene and the effective model is hard to be constructed.
\end{abstract}

\maketitle

%\section{Introduction} 
\emph{Introduction.}---Twisted bilayer graphene (TBG) has recently been intensely researched in both theory and experiment \cite{PhysRevLett.99.256802,Bistritzer12233,Li2010,PhysRevLett.107.216602,PhysRevLett.108.216802,Maher2013,Hunt2017f}. When the twisted angle ($\theta$) is decreased from large value down to the so-called first ``magic angle" ($\theta=1.08^\circ$), the system starts exhibiting correlated insulating phases \cite{Cao201882,Sharpe605,Choi2019,Kerelsky2019,Lu2019,Codecidoeaaw9770} and unusual superconductivity \cite{Cao201843,Yankowitz1059,Lu2019,Codecidoeaaw9770} at special fillings in experiments. It is widely believed that these novel transport properties originate from the nearly flat bands (FBs) that develop near the charge neutrality point (Fermi energy of TBG) at the magic angle \cite{Bistritzer12233}. The measured bandwidth of the FBs is in the order of 10-20 meV from tunneling spectroscopy experiments \cite{Choi:2019aa,Kerelsky2019,Xie:2019aa,Jiang:2019aa}. Angle-resolved photoemission spectroscopy measurements \cite{Lisi2020} find that the FBs are separated by a gap of 30-50 meV from both higher and lower energy ``passive" bands. Although theoretical effective models have been shown to capture the essential properties of the FBs in TBG, the parameters of effective models are estimated according to experimental results which often include the correlation effects of Coulomb interaction. It is not easy to get the accurate electronic structures purely from $ab~initio$ calculation without involving any approximation for small angle cases. Untangling these correlations from the underlying single-particle band structure upon which they act becomes difficult and $ab~initio$ studies of this problem are desirable.

To solve this problem, researchers attempted to perform $ab~initio$ calculations within density functional theory (DFT) for TBG. 
Uchida \emph{et~al}. \cite{PhysRevB.90.155451} carried out large-scale DFT calculations for structural relaxation and band structure within a real-space scheme. They found the Fermi velocity of the Dirac electron dramatically reduced toward zero when $\theta <5^\circ$, and it vanished at $\theta=1.08^\circ$ and then slightly increased for the smaller $\theta$. 
Song \emph{et~al}. \cite{PhysRevLett.123.036401} performed single-particle DFT calculations for rigid TBG using the Vienna Ab initio Simulation Package (VASP) based on plane-wave pseudo-potential method. They calculated the eigenenergies only at $\Gamma$ and $K$ momenta in the moir\'{e} Brillouin zone (MBZ) when $\theta \leqslant 1.08^\circ$, while the whole band structure along high symmetrical lines for $\theta =1.41^\circ$ and larger ones. 
Recently, Cantele \emph{et~al}. \cite{PhysRevB.99.195419,PhysRevResearch.2.043127} obtained the relaxed TBG structures and the bands at the magic angle and other larger ones by VASP. Their results show that the FBs at the first magic angle are connected with the atomic displacements originating from the interlayer van der Waals interaction, and the main contribution to the experimental gaps at the $\Gamma$ point is the out-of-plane displacements.
The computational cost of all the aforementioned approaches is in order of $N^3$, namely $O(N^3)$, where $N$ is the number of atoms per unit cell, and $N$ is more than tens of thousand when $\theta$ is around the first magic angle. The further study of TBG with smaller angles within such DFT calculation is unaffordable.

In this paper, we develop a two-step DFT calculation workflow which can dramatically reduce the computational cost while achieve satisfied accuracy. We perform this two-step calculation for both rigid and corrugated TBG down to the first magic angle. The computational cost is of $O(N)$, which is much reduced comparing with that of VASP. The band structure, bandwidth of FBs and the band gap at $K$ point have been shown and discussed. They are in good agreement with those obtained from tight-binding (TB) models, continuum models, VASP calculations, as well as experimental observations. This method can be applied on TBG with even smaller angles and other twisted two-dimensional materials, whose effective models are hard to be constructed, such as twisted trilayer graphene, twisted double bilayer graphene, twisted bilayer transition-metal dichalcogenides.

%\subsection{Workflow of calculations}
\begin{figure}
\includegraphics[width=1\columnwidth]{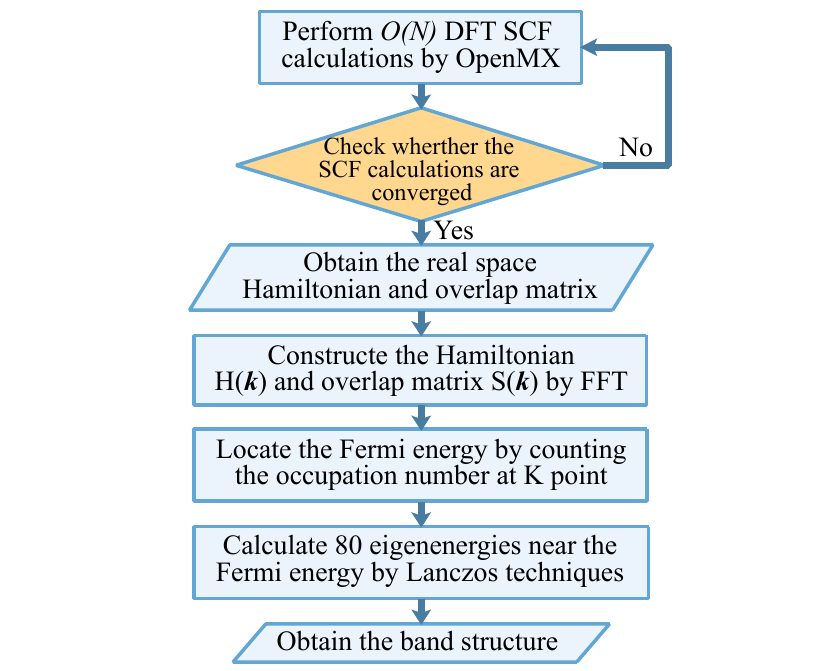}
\caption{(Color online) Calculation workflow contains the SCF calculation by $O(N)$ Krylov subspace method implemented in OpenMX and band structure calculation with shift-invert and Lanczos techniques.} 
\label{fig:1}
\end{figure}

\emph{Workflow of the method.}---The workflow is shown in Fig.~\ref{fig:1}. The first step is to perform self-consistent field (SCF) calculation within DFT using the $O(N)$ method implemented in OpenMX \cite{PhysRevB.67.155108}, which is an $ab~initio$ software package based on localized pseudo-atomic orbitals (PAOs) and pseudo-potential method. The details about the crystal structure of rigid and corrugated TBG and the calculation method are introduced in the Supplemental Material \cite{SupplementalMaterial}. The convergence criterion is set as the difference in total energies of the last two SCF loops being smaller than $10^{-7}$ Hartree. If the SCF calculation is not converged within 200 loops, the parameters controlling the mixing of charge densities in previous loops will be modified and restart the SCF calculation. If the SCF calculation is converged, we obtain the real space Hamiltonian and the overlap matrix both in PAO basis set. 
Then, the momentum $\boldsymbol{k}$ dependent Hamiltonian $H(\boldsymbol{k})$ and overlap matrix $S(\boldsymbol{k})$ can be constructed using fast Fourier transformation (FFT), which is very similar to the interpolation method based on Wannier functions constructed from DFT band calculation. The dimension of $H(\boldsymbol{k})$ and $S(\boldsymbol{k})$ is quite big in small angle case, and to solve its eigenvalues is very challenge. Due to the localized PAO basis set and the cutoff of real space hopping interaction among electrons, $H(\boldsymbol{k})$ and $S(\boldsymbol{k})$ are sparse matrices. Lanczos technique combined with the shift-invert spectral transformation is suitable for solving only a small number of eigenvalues (out of the large number of eigenvalues at small angles) near a selected value of energy, which is much faster than direct full diagonalization for all eigenvalues. The details about this techniques are introduced in the Supplemental Material \cite{SupplementalMaterial}. The Fermi energy is located by counting the occupation number of the bands at $K$ by direct diagonalization. Finally, eighty eigenenergies near the Fermi level are calculated at a series of momenta in the MBZ by shift-invert and Lanczos techniques and thus the band structures can been obtained. 
We have also implemented ScaLAPACK library support for calculation, allowing us to obtain the full spectrum down $\theta=1.41^\circ$ providing the other consistency check.

%\subsection{$O(N)$ methods}
\begin{figure}
\includegraphics[width=1\columnwidth]{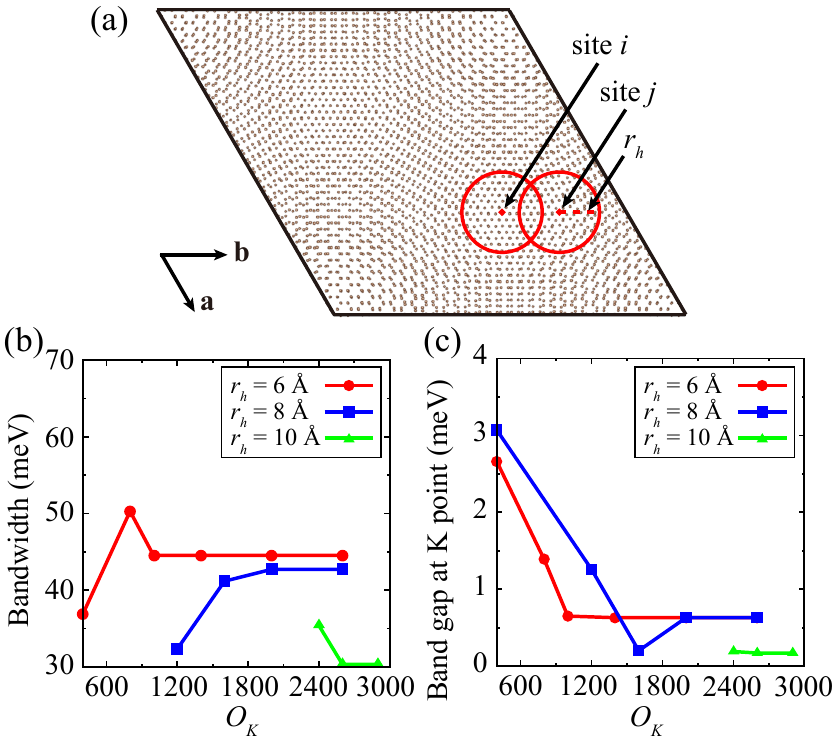}
\caption{(Color online) The top view of crystal structure of TBG with $i$ = 16, $\theta=2.00^\circ$. It contains 3268 atoms and each carbon atom site has thirteen localized PAOs as basis set. Two red circles represent truncated clusters centered on C atoms at sites $i$ and $j$ with $r_h$ = 8 \AA. Bandwidth of FBs (b) and band gap at $K$ point (c) obtained with different parameters of hopping range $r_h$ and dimension of Krylov subspace $O_K$ for rigid TBG at $\theta=1.41^\circ$.
\label{fig:2}}
\end{figure}

\emph{$O(N)$ method.}---The conventional DFT methods are not efficient for large systems like TBG when $\theta$ is close to magic angle. We employ the Krylov subspace $O(N)$ method implemented in OpenMX to perform the SCF calculation. This method has been well explained in Ref.~\onlinecite{PhysRevB.74.245101}, and we just briefly introduce the parts closely related to our calculation. For a large system, the atoms inside a sphere centered on each site are physically truncated into small clusters, as shown in Fig.~\ref{fig:2} (a). The radius of the sphere is so-called hopping range $r_h$. If the distance between the neighboring sites and the central atom is less than $r_h$, the hopping of electrons between them is considered. For each truncated cluster, the vector space $\mathbf{U}_{C}$ defined by basis functions in the truncated cluster is mapped to a Krylov subspace $\mathbf{U}_{K}$. In general, the dimension of $\mathbf{U}_{K}$ ($O_K$) is smaller than that of $\mathbf{U}_{C}$, and the dimension of $\mathbf{U}_{C}$ is much smaller than that of total vector space $\mathbf{U}_{F}$ spanned by all of the basis functions in the whole unit cell. Thus, the computational cost is significantly reduced compared to the conventional DFT methods.

The key input parameters are $r_h$ and $O_K$, which determine the balance between accuracy and computational cost. To set their values properly, we test them for rigid TBG at $\theta=1.41^\circ$ using this two-step method. The bandwidth of FBs and the band gap at $K$ obtained from different $r_h$ and $O_K$ have been shown in Figs.~\ref{fig:2} (b) and \ref{fig:2} (c). For $r_h$ = 6, 8 and 10 \AA, the truncated cluster contains 84 atoms, 168 atoms and 222 atoms, whose dimension of Hamiltonian matrix is 1092, 2184, 2886, respectively, and the results are converged when $O_K$ = 1000, 2000 and 2600, respectively, which are about 8.9\% smaller than the dimension of the Hamiltonian matrix ($\mathbf{U}_{C}$) on avarage. However, it is noted that the cutoff radius of PAOs for basis function of C atom is 3.18 \AA, and the maximum distance between two C sites with their PAOs overlapped is 6.36 \AA. The $r_h$ = 6 \AA~is too small for the truncated clusters. On the other hand, $r_h$ = 10 \AA~is too large, which includes too many useless non-orthogonal basis functions without any hopping interaction with the central C atom in the truncated clusters. This leads to the over completeness of the basis set and gives out quite strange results. Therefore, $r_h$ = 8 \AA~is quite reasonable choice. In fact, the band structures with these three $r_h$ are somehow different from each other. 
We check these bands against those from TB models, continuum models and VASP calculations. The most reliable one is from calculations with $r_h$ = 8 \AA. Because the cutoff radius of PAOs and the total number of atoms in a truncated cluster with the same $r_h$ hardly change at different $\theta$, the appropriate $r_h$ and $O_K$ should be independent of $\theta$. This is verified by our further tests for both rigid and corrugated structures at different $\theta$. Hence, we use $r_h$ = 8 \AA~and $O_K$ = 2000 in calculations for all TBG structures.

%\subsection{Calculation efficient}
\begin{figure}
\includegraphics[width=1\columnwidth]{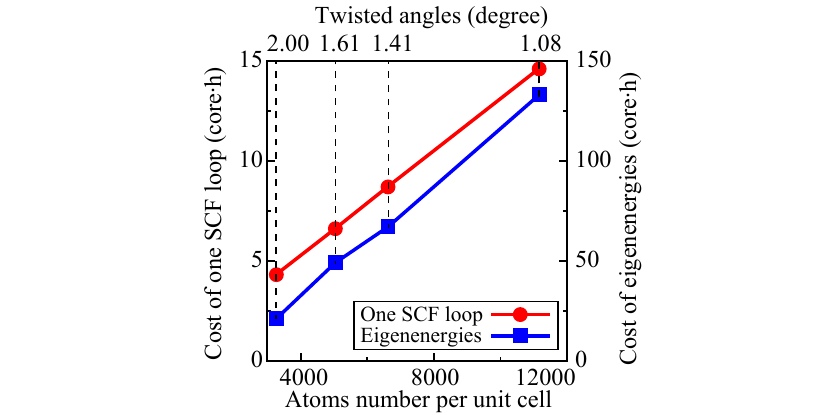}
\caption{(Color online) The computational cost of one SCF loop (red dots) and selected eighty eigenenergies solved at one momentum (blue squares) for rigid TBG of four twisted angles. The unit cell contains 3268, 5044, 6628 and 11164 carbon atoms at $\theta=2.00^\circ$, $1.61^\circ$, $1.41^\circ$ and $1.08^\circ$, respectively.
\label{fig:3}}
\end{figure}

\emph{Calculation efficiency.}---We show the computational cost of rigid TBGs at four different $\theta$ values in Fig.~\ref{fig:3}. These calculations are performed with Intel Xeon CPU E3-1240 v6 family at base frequency of 3.7 GHz. By using $O(N)$ method, the computational cost, in unit of core$\cdot$hour, of one SCF loop calculation almost linearly increases with the number of atoms per unit cell. Because TBG contains tens of thousands of atoms in a unit cell at small angles, the $O(N)$ method is considerably faster than the conventional DFT methods of $O(N^3)$. The cost for obtaining 80 eigenenergies near Fermi energy at one momentum is also nearly $O(N)$, because the dimensions of Hamiltonian matrix and overlap matrix are of $O(N)$. 
In Fig.~\ref{fig:3}, one SCF loop takes about 4.3 core$\cdot$h, 6.6 core$\cdot$h, 8.7 core$\cdot$h, 14.6 core$\cdot$h, and one band calculation at one momentum point needs about 21 core$\cdot$h, 49 core$\cdot$h, 67 core$\cdot$h, 133 core$\cdot$h at $\theta=2.00^\circ$, $\theta=1.61^\circ$, $\theta=1.41^\circ$, $\theta=1.08^\circ$, respectively. We also get the similar results for  corrugated structures. As a reference, Ref.~\onlinecite{PhysRevLett.123.036401} stated that the cost of about 576 core$\cdot$h, 1728 core$\cdot$h and 17280 core$\cdot$h is needed to finish one band calculation at one momentum point by VASP at $\theta=2.00^\circ$, $\theta=1.41^\circ$ and $\theta=1.08^\circ$, respectively. Our band calculation is about 130 times faster than that of VASP at $\theta=1.08^\circ$. It is also noted that the total computational cost by VASP requires $2.1\times10^6$ core$\cdot$h to $4.1\times10^6$ core$\cdot$h including one relaxation calculation, one SCF calculation and band calculations at seven momentum points of TBG at $\theta=1.08^\circ$ \cite{PhysRevB.99.195419,PhysRevResearch.2.043127}.

%\section{Band results}
\begin{figure*}
\includegraphics[width=2\columnwidth]{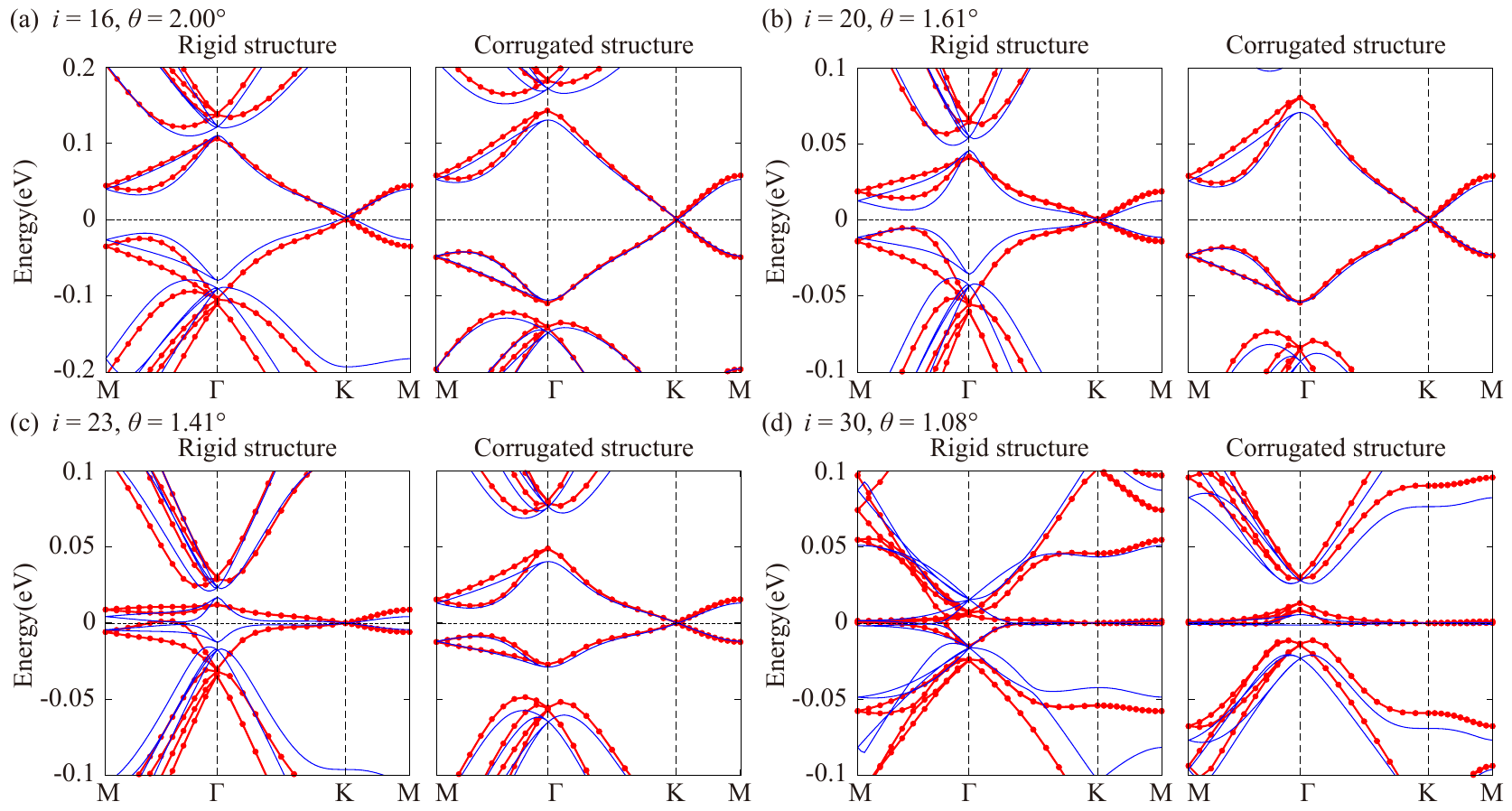}
\caption{The band structures obtained from our two-step method (red dots) and TB model (blue lines) for rigid (left) and corrugated (right) TBGs at twisted angles: (a) $\theta=2.00^\circ$, (b) $\theta=1.61^\circ$, (c) $\theta=1.41^\circ$ and (d) $\theta=1.08^\circ$.
\label{fig:4}}
\end{figure*}

\begin{figure}
\includegraphics[width=1\columnwidth]{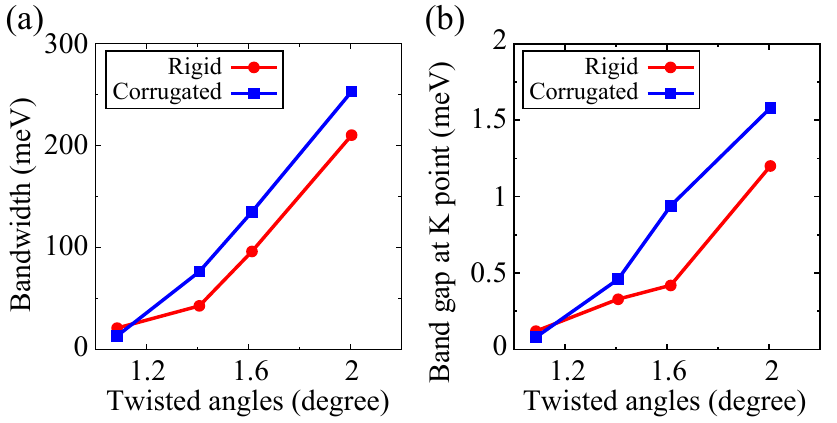}
\caption{(Color online) Bandwidth of the four bands near Fermi level (a) and band gap at $K$ point (b) for rigid (red dots) and corrugated (blue squares) TBGs at four twisted angles: $\theta=2.00^\circ$, $1.61^\circ$, $1.41^\circ$, and $1.08^\circ$.
\label{fig:5}}
\end{figure}

\emph{Results of the method.}---The band structures for rigid and corrugated TBGs at four twist angles are shown in Fig. \ref{fig:4}. The red points represent the bands calculated by the two-step method, and the red lines are plotted as guide to the eye. The blue lines are bands obtained by TB models \cite{PhysRevB.99.155415}. They are in quite good agreement, as well as with those from continuum models \cite{Bistritzer12233,PhysRevB.87.205404}. We firstly focus on the bands of rigid structures. The bandwidth of the four bands (two per valley in the continuum models) near the Fermi energy monotonically decreases with $\theta$ as shown in Fig.~\ref{fig:5} (a). When $\theta=1.08^\circ$, the bandwidth decreases to 21.0 meV and the FBs appear at the Fermi energy (see Fig.~\ref{fig:4}(d)). They are separated from the lower energy bands by a gap of 8.4 meV. This is a little different from those calculated from both TB model and continuum model where the FBs entangled with the lower energy bands at $\Gamma$. The other difference is that there exists a gap among these four bands at $K$ point from this method and VASP \cite{PhysRevLett.123.036401}, because the inter-valley coupling in TBG is not ignored in DFT calculation, while it is four-fold degenerate at $K$ point in model calculations where the inter-valley coupling is ignored. This gap gradually decreases with $\theta$ and becomes very small at the first magic angle (0.12 meV) of $\theta=1.08^\circ$ as shown in Fig.~\ref{fig:5} (b).

Next, we discuss the results for corrugated structures. The corrugation enlarges both the bandwidth and the gap at $K$ point when $\theta>1.08^\circ$, but it reduces them at the magic angle $\theta=1.08^\circ$ (see Fig.~\ref{fig:5}). We focus on the band structure at $\theta=1.08^\circ$. The four bands become FBs with bandwidth of about 13.2 meV, and the gap at $K$ point decreases to 0.08 meV. The two gaps separating FBs from the higher and lower energy bands increase due to the corrugations \cite{PhysRevX.8.031087,PhysRevB.99.155415}, being 15.3 meV and 11.3 meV, respectively. These results are consistent with those from TB models \cite{PhysRevB.99.155415,PhysRevX.8.031087,PhysRevB.96.075311,PhysRevB.98.085144,2019arXiv191012805L,PhysRevB.98.235137,PhysRevB.92.075402}, continuum models \cite{PhysRevB.99.155415,PhysRevResearch.1.013001}, VASP calculations with fully relaxed structures \cite{PhysRevB.99.195419,PhysRevResearch.2.043127} and experimental observations \cite{Cao201882,Cao201843,Kerelsky2019,Xie:2019aa,Jiang:2019aa,Choi:2019aa,Lisi2020}. This implies that atomic corrugations in TBG play an important role in the electronic structures, and that the main effects of corrugations in real materials can be simulated by modeled corrugations.

%\section{Conclusions}
\emph{Conclusions.}---We design a two-step workflow to carry out DFT calculation of band structures for large systems like TBGs with tens of thousands of atoms. In the first step, a SCF calculation using $O(N)$ Krylov subspace method implemented in OpenMX is performed to obtain the Hamiltonian and overlap matrix in real space. In the second step, the momentum dependent Bloch Hamiltonian and overlap matrix are constructed in the sense of interpolation, and only a small number of eigenenergies near the Fermi energy are solved by shift-invert and Lanczos techniques. The key input parameters in $O(N)$ SCF calculation are $r_h$ and $O_K$, which can be systematically tuned to balance the accuracy and computational cost. We find that $r_h$ = 8 \AA~and $O_K$ = 2000 are suitable for TBG and that these values do not depend much on $\theta$ and corrugation. The computational cost linearly increases with the number of atoms per unit cell for both rigid TBGs and those with corrugation. For rigid TBG at $\theta=1.08^\circ$, one SCF loop requires about 14.6 core$\cdot$h and one band calculation of 80 eigenvalues at one momentum point requires about 133 core$\cdot$h on our computers. This is affordable for even larger system with smaller angles. The band structures obtained at four twisted angles by this method are consistent with those of TB models, continuum models, VASP calculations and the experimental observations. This method can be applied to other twisted materials and could become an efficient tool when no effective model is available.

We acknowledge the supports from the National Natural Science Foundation (Grant No. 11925408), the Ministry of Science and Technology of China (Grants No. 2016YFA0300600 and 2018YFA0305700), the Chinese Academy of Sciences (Grant No. XDB33000000), the K. C. Wong Education Foundation (GJTD-2018-01), the Beijing Natural Science Foundation (Z180008), and the Beijing Municipal Science and Technology Commission (Z191100007219013). B. A. B and N. R. were also supported by the DOE Grant No. DE-SC0016239, the Schmidt Fund for Innovative Research, Simons Investigator Grant No. 404513, the Packard Foundation, the Gordon and Betty Moore Foundation through Grant No. GBMF8685 towards the Princeton theory program, and a Guggenheim Fellowship from the John Simon Guggenheim Memorial Foundation. Further support was provided by the NSF-EAGER No. DMR 1643312, NSF-MRSEC No. DMR-1420541 and DMR-2011750, ONR No. N00014-20-1-2303, Gordon and Betty Moore Foundation through Grant GBMF8685 towards the Princeton theory program, BSF Israel US foundation No. 2018226, and the Princeton Global Network Funds.

\bibliography{TBGpaper}
\end{document}